\documentclass[11pt, a4paper]{article}\usepackage[]{graphicx}\usepackage[]{color}
\makeatletter
\def\maxwidth{ %
  \ifdim\Gin@nat@width>\linewidth
    \linewidth
  \else
    \Gin@nat@width
  \fi
}
\makeatother

\definecolor{fgcolor}{rgb}{0.345, 0.345, 0.345}
\newcommand{\hlnum}[1]{\textcolor[rgb]{0.686,0.059,0.569}{#1}}%
\newcommand{\hlstr}[1]{\textcolor[rgb]{0.192,0.494,0.8}{#1}}%
\newcommand{\hlcom}[1]{\textcolor[rgb]{0.678,0.584,0.686}{\textit{#1}}}%
\newcommand{\hlopt}[1]{\textcolor[rgb]{0,0,0}{#1}}%
\newcommand{\hlstd}[1]{\textcolor[rgb]{0.345,0.345,0.345}{#1}}%
\newcommand{\hlkwa}[1]{\textcolor[rgb]{0.161,0.373,0.58}{\textbf{#1}}}%
\newcommand{\hlkwb}[1]{\textcolor[rgb]{0.69,0.353,0.396}{#1}}%
\newcommand{\hlkwc}[1]{\textcolor[rgb]{0.333,0.667,0.333}{#1}}%
\newcommand{\hlkwd}[1]{\textcolor[rgb]{0.737,0.353,0.396}{\textbf{#1}}}%

\usepackage{framed}
\makeatletter
\newenvironment{kframe}{%
 \def\at@end@of@kframe{}%
 \ifinner\ifhmode%
  \def\at@end@of@kframe{\end{minipage}}%
  \begin{minipage}{\columnwidth}%
 \fi\fi%
 \def\FrameCommand##1{\hskip\@totalleftmargin \hskip-\fboxsep
 \colorbox{shadecolor}{##1}\hskip-\fboxsep
     \hskip-\linewidth \hskip-\@totalleftmargin \hskip\columnwidth}%
 \MakeFramed {\advance\hsize-\width
   \@totalleftmargin\z@ \linewidth\hsize
   \@setminipage}}%
 {\par\unskip\endMakeFramed%
 \at@end@of@kframe}
\makeatother

\definecolor{shadecolor}{rgb}{.97, .97, .97}
\definecolor{messagecolor}{rgb}{0, 0, 0}
\definecolor{warningcolor}{rgb}{1, 0, 1}
\definecolor{errorcolor}{rgb}{1, 0, 0}
\newenvironment{knitrout}{}{} 

\usepackage{alltt}

\usepackage[english]{babel}
\usepackage[utf8]{inputenc}
\usepackage{hyperref}
\usepackage{amsmath}
\usepackage{amssymb}
\usepackage{parskip}
\usepackage{a4wide}
\usepackage{natbib}
\usepackage{hyperref}
\usepackage[bf]{caption}
\usepackage{float}

\usepackage{mathtools}
\mathtoolsset{showonlyrefs}

\author{
  Mikkel Meyer Andersen\\
  \url{mikl@math.aau.dk} \\
  Department of Mathematical Sciences \\
  Aalborg University \\
  Denmark
  \and
  Poul Svante Eriksen\\
  \url{svante@math.aau.dk} \\
  Department of Mathematical Sciences \\
  Aalborg University \\
  Denmark
  \and
  Niels Morling\\
  \url{niels.morling@sund.ku.dk} \\
  Section of Forensic Genetics \\
  Department of Forensic Medicine \\
  Faculty of Health and Medical Sciences \\
  University of Copenhagen \\
  Denmark
}

\title{A gentle introduction to the discrete Laplace method for estimating Y-STR haplotype frequencies}
\date{}
\IfFileExists{upquote.sty}{\usepackage{upquote}}{}

\begin{document}
\maketitle

\begin{abstract}
Y-STR data simulated under a Fisher-Wright model of evolution with a single-step mutation model turns out to be well predicted by a method using discrete Laplace distributions.
\end{abstract}

\tableofcontents

\section{Introduction}
This tutorial introduces the discrete Laplace method for estimating Y-STR haplotype frequencies as described by \cite{AndersenDisclap2013}.

To accomplish this, we demonstrate a number of examples using \texttt{R} \citep{R}. The code examples look like the following that loads the \texttt{disclap} package \citep{disclap14} which is needed for the following examples:
\begin{knitrout}
\definecolor{shadecolor}{rgb}{0.933, 0.933, 0.933}\color{fgcolor}\begin{kframe}
\begin{alltt}
\hlkwd{library}\hlstd{(disclap)}
\end{alltt}
\end{kframe}
\end{knitrout}

If you do not have installed the \texttt{disclap} package, please visit \url{http://cran.r-project.org/package=disclap}.

\section{The discrete Laplace distribution}
The discrete Laplace distribution is a probability distribution like e.g.\ the binomial distribution or the normal/Gaussian distribution.

The discrete Laplace distribution has two parameters: a dispersion parameter $0 < p < 1$ and a location parameter $y \in \mathbb{Z} = \{\ldots, -2, -1, 0, 1, 2, \ldots \}$.

Let $X \sim DL(p, y)$ denote that the random variable $X$ follows a discrete Laplace distribution with dispersion parameter $0 < p < 1$ and location parameter $y$. Then a realisation of the random variable, $X = x$, can be any integer in $\mathbb{Z}$. The random variable $X$ has the probability mass function given by
\begin{align}
  f(X = x; p, y) = \frac{1-p}{1+p} \cdot p^{\vert x - y \vert} \quad \text{for $x \in \mathbb{Z}$}.
\end{align}

As seen, only the absolute value of $x - y$ is used. This means that the probability mass function is symmetric around $y$.

Let us try to plot the probability mass function $f(X = x; p, y)$ for $p=0.3$ and $y=13$ from $x=8$ to $x=18$:
\begin{figure}[H]
\begin{center}
\begin{knitrout}
\definecolor{shadecolor}{rgb}{0.933, 0.933, 0.933}\color{fgcolor}\begin{kframe}
\begin{alltt}
\hlstd{p} \hlkwb{<-} \hlnum{0.3}
\hlstd{y} \hlkwb{<-} \hlnum{13}
\hlstd{x} \hlkwb{<-} \hlkwd{seq}\hlstd{(}\hlnum{8}\hlstd{,} \hlnum{18}\hlstd{,} \hlkwc{by} \hlstd{=} \hlnum{1}\hlstd{)}
\hlkwd{barplot}\hlstd{(}\hlkwd{ddisclap}\hlstd{(x} \hlopt{-} \hlstd{y, p),} \hlkwc{names} \hlstd{= x,} \hlkwc{xlab} \hlstd{=} \hlstr{"x, e.g. Y-STR allele"}\hlstd{,}
    \hlkwc{ylab} \hlstd{=} \hlkwd{paste}\hlstd{(}\hlstr{"Probability mass, f(X = x; "}\hlstd{, p,} \hlstr{", "}\hlstd{, y,} \hlstr{")"}\hlstd{,} \hlkwc{sep} \hlstd{=} \hlstr{""}\hlstd{))}
\end{alltt}
\end{kframe}
\includegraphics[width=12cm]{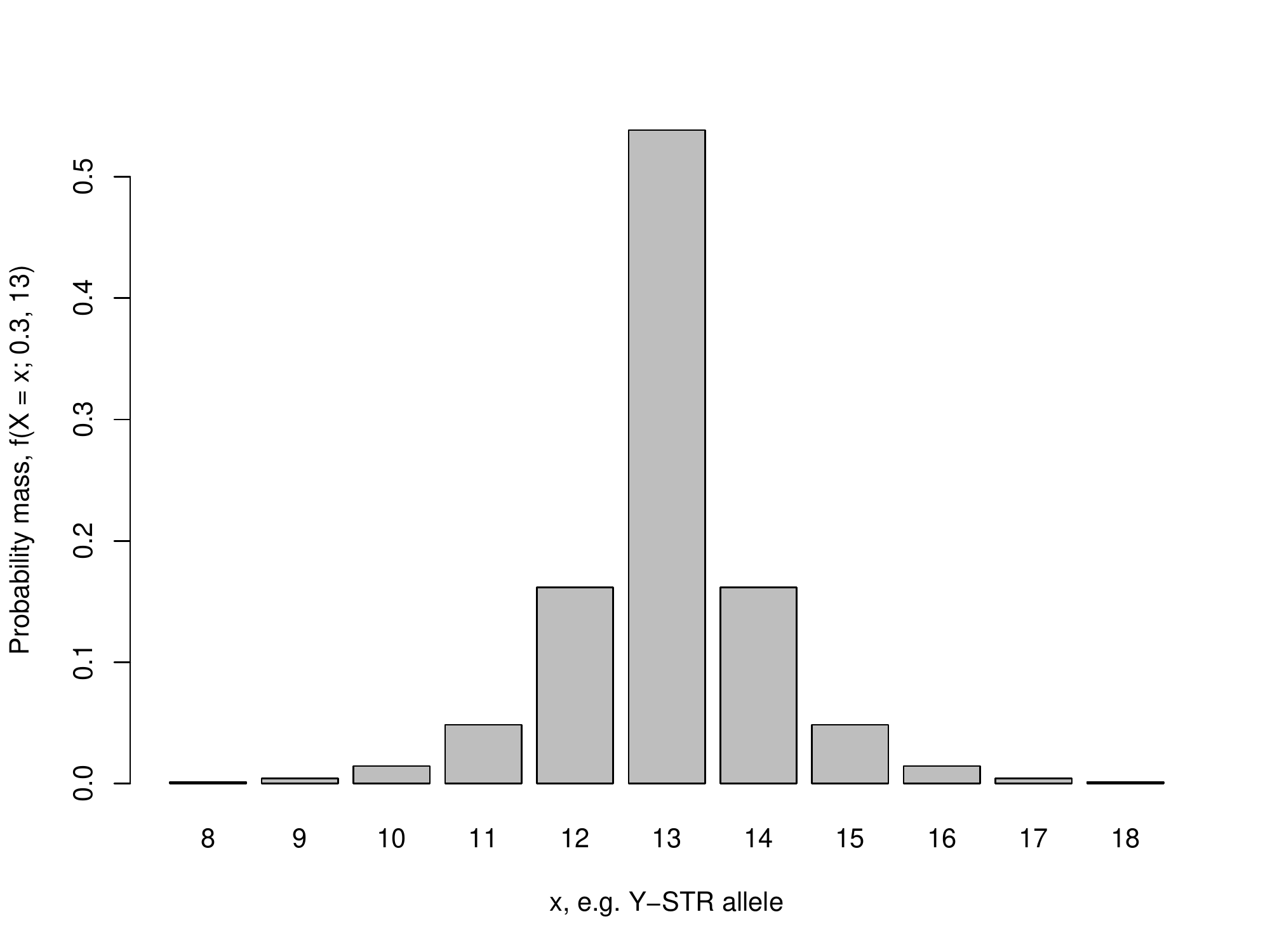} 

\end{knitrout}

\caption{The probability mass function, $f(X = x; p, y)$, for the discrete Laplace distribution with dispersion parameter $p=0.3$ and location parameter $y=13$ from $x=8$ to $x=18$.}
\end{center}
\end{figure}

We plot the distribution for values of $x$ from $8$ to $18$ as there is almost no probability mass outside these values. We can find out how much of the probability mass that we have plotted:
\begin{knitrout}
\definecolor{shadecolor}{rgb}{0.933, 0.933, 0.933}\color{fgcolor}\begin{kframe}
\begin{alltt}
\hlkwd{sum}\hlstd{(}\hlkwd{ddisclap}\hlstd{(}\hlkwd{abs}\hlstd{(x} \hlopt{-} \hlstd{y), p))}
\end{alltt}
\begin{verbatim}
## [1] 0.9989
\end{verbatim}
\end{kframe}
\end{knitrout}

Thus, only 0.0011 of the probability mass is outside $\{8, 9, \ldots, 17, 18\}$.

If we have a sample of realisations from $X \sim DL(p, y)$ denoted by $\{ x_i \}_{i=1}^n$, then maximum likelihood estimates are given by the following quantities \citep{AndersenDisclap2013}:
\begin{align}
  \hat{y} &= \text{median} \{ x_i \}_{i=1}^n , \\
  \hat{\mu} &= \frac{1}{n} \sum_{i=1}^n \vert x_i - \hat{y} \vert \, \text{and} \\
  \hat{p} &= \hat{\mu}^{-1} \left (\sqrt{\hat{\mu}^2 + 1} - 1 \right ) .
\end{align}

\newpage
Example:
\begin{knitrout}
\definecolor{shadecolor}{rgb}{0.933, 0.933, 0.933}\color{fgcolor}\begin{kframe}
\begin{alltt}
\hlkwd{set.seed}\hlstd{(}\hlnum{1}\hlstd{)}  \hlcom{# Makes it possible to reproduce the simulation results}
\hlstd{p} \hlkwb{<-} \hlnum{0.3}  \hlcom{# Dispersion parameter }
\hlstd{y} \hlkwb{<-} \hlnum{13}  \hlcom{# Location parameter }
\hlstd{x} \hlkwb{<-} \hlkwd{rdisclap}\hlstd{(}\hlnum{100}\hlstd{, p)} \hlopt{+} \hlstd{y}  \hlcom{# Generate a sample using the rdisclap function}
\hlstd{y.hat} \hlkwb{<-} \hlkwd{median}\hlstd{(x)}
\hlstd{y.hat}
\end{alltt}
\begin{verbatim}
## [1] 13
\end{verbatim}
\begin{alltt}
\hlstd{mu.hat} \hlkwb{<-} \hlkwd{mean}\hlstd{(}\hlkwd{abs}\hlstd{(x} \hlopt{-} \hlstd{y.hat))}
\hlstd{mu.hat}
\end{alltt}
\begin{verbatim}
## [1] 0.57
\end{verbatim}
\begin{alltt}
\hlstd{p.hat} \hlkwb{<-} \hlstd{mu.hat}\hlopt{^}\hlstd{(}\hlopt{-}\hlnum{1}\hlstd{)} \hlopt{*} \hlstd{(}\hlkwd{sqrt}\hlstd{(mu.hat}\hlopt{^}\hlnum{2} \hlopt{+} \hlnum{1}\hlstd{)} \hlopt{-} \hlnum{1}\hlstd{)}
\hlstd{p.hat}  \hlcom{# We expect 0.3}
\end{alltt}
\begin{verbatim}
## [1] 0.265
\end{verbatim}
\begin{alltt}
\hlcom{# The observed distribution of d's}
\hlstd{tab} \hlkwb{<-} \hlkwd{prop.table}\hlstd{(}\hlkwd{table}\hlstd{(x))}
\hlstd{tab}
\end{alltt}
\begin{verbatim}
## x
##   10   11   12   13   14   15   16 
## 0.01 0.03 0.15 0.55 0.20 0.05 0.01
\end{verbatim}
\end{kframe}
\end{knitrout}

\newpage
This can be plotted against the expected counts as follows:
\begin{figure}[H]
\begin{center}
\begin{knitrout}
\definecolor{shadecolor}{rgb}{0.933, 0.933, 0.933}\color{fgcolor}\begin{kframe}
\begin{alltt}
\hlkwd{plot}\hlstd{(}\hlnum{1}\hlopt{:}\hlkwd{length}\hlstd{(tab),} \hlkwd{ddisclap}\hlstd{(}\hlkwd{as.integer}\hlstd{(}\hlkwd{names}\hlstd{(tab))} \hlopt{-} \hlstd{y.hat, p.hat),}
  \hlkwc{type} \hlstd{=} \hlstr{"h"}\hlstd{,} \hlkwc{col} \hlstd{=} \hlstr{"#999999"}\hlstd{,} \hlkwc{lend} \hlstd{=} \hlstr{"butt"}\hlstd{,} \hlkwc{lwd} \hlstd{=} \hlnum{50}\hlstd{,}
  \hlkwc{xlab} \hlstd{=} \hlstr{"x, e.g. Y-STR allele"}\hlstd{,} \hlkwc{ylab} \hlstd{=} \hlstr{"Probability mass"}\hlstd{,} \hlkwc{axes} \hlstd{=} \hlnum{FALSE}\hlstd{)}
\hlkwd{axis}\hlstd{(}\hlnum{1}\hlstd{,} \hlkwc{at} \hlstd{=} \hlnum{1}\hlopt{:}\hlkwd{length}\hlstd{(tab),} \hlkwc{labels} \hlstd{=} \hlkwd{names}\hlstd{(tab))}
\hlkwd{axis}\hlstd{(}\hlnum{2}\hlstd{)}
\hlkwd{points}\hlstd{(}\hlnum{1}\hlopt{:}\hlkwd{length}\hlstd{(tab), tab,} \hlkwc{type} \hlstd{=} \hlstr{"h"}\hlstd{,} \hlkwc{col} \hlstd{=} \hlstr{"#000000"}\hlstd{,}
  \hlkwc{lend} \hlstd{=} \hlstr{"butt"}\hlstd{,} \hlkwc{lwd} \hlstd{=} \hlnum{25}\hlstd{)}
\hlkwd{legend}\hlstd{(}\hlstr{"topright"}\hlstd{,} \hlkwd{c}\hlstd{(}\hlstr{"Estimated distribution"}\hlstd{,} \hlstr{"Observations"}\hlstd{),}
  \hlkwc{pch} \hlstd{=} \hlnum{15}\hlstd{,} \hlkwc{col} \hlstd{=} \hlkwd{c}\hlstd{(}\hlstr{"#999999"}\hlstd{,} \hlstr{"#000000"}\hlstd{))}
\end{alltt}
\end{kframe}
\includegraphics[width=12cm]{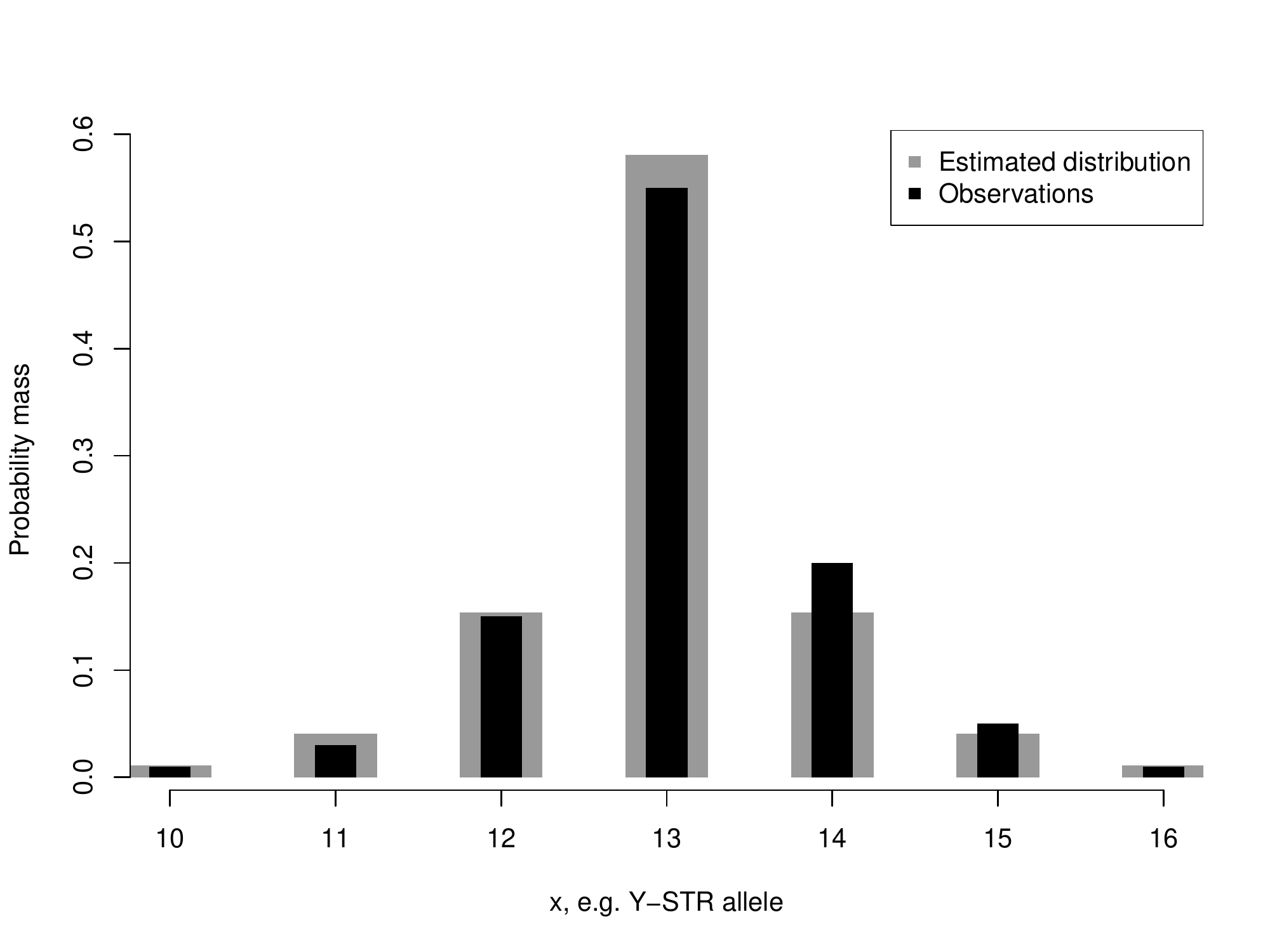} 

\end{knitrout}

\caption{Observed frequencies of the $x$'s compared to a discrete Laplace distribution with parameters estimated from the sample.}
\label{fig:simple-example} 
\end{center}
\end{figure}

\section{Mixtures of multivariate, marginally independent, discrete Laplace distributions}
Assume a very simple 'haplotype' with only one locus. Also assume a simple and isolated population. Then, it is reasonable to assume that there is a modal/central Y-STR allele, $y$, and that all the alleles are distributed around this allele.

If we go back to Figure~\ref{fig:simple-example}, this can be illustrated by $y=13$ as the central Y-STR allele and a distribution around $y=13$ with shorter and longer alleles. 

To begin with, it might seem a bit overwhelming that Y-STR alleles should follow a simple probabiity distribution such as the discrete Laplace distribution. But surprisingly, it is actually a good approximation as demonstrated by \cite{AndersenDisclap2013}.

We have haplotypes with several loci. When we assess multiple loci haplotypes, we assume that mutations happen independently across loci. Each locus has its own discrete Laplace distribution of allele probabilities, and the probability of a haplotype is the product of probabilities across loci. This gives a multivariate discrete Laplace distribution, where the marginals (that is, at each locus) are independent, discrete Laplace distributions.

Just as before, for a one locus haplotype, we can assume that there is a modal/central Y-STR profile with $r$ loci, $y=(y_1, y_2, \ldots, y_r)$, and all the alleles are distributed around this profile. We also assume that the discrete Laplace distribution at each locus has its own parameter, where $p_k$ is the parameter at the $k$\textsuperscript{th} locus. Normally, the central Y-STR profile, $y$, would also be regarded as parameters.

As before, let $f(x; p, y)$ be the probability mass function of a discrete Laplace distribution. We define an observation $X = (X_{1}, X_{2}, \ldots, X_{r})$ to be from a multivariate distribution of independent, discrete Laplace distributions when the probability of observing $X=x$ is
\begin{align} \label{eq:disclap-independent}
  \prod_{k=1}^r f \left ( x_{k} ; p_{k}, y_{k} \right ).
\end{align}
This corresponds to that the individual $X$ has mutated away from $y$ independently at each locus. 

Now, we have one more generalisation. A population may have several subpopulations, e.g.\ introduced by migration or by evolution. This means that we need to have a mixture of multivariate distributions with marginally independent, discrete Laplace distributions. Each component in the mixture represents a subpopulation. We define an observation $X = (X_{1}, X_{2}, \ldots, X_{r})$ to be from a mixture of multivariate, marginally independent, discrete Laplace distributions, when the probability of observing $X=x$ is
\begin{align} \label{eq:disclap-mixtures}
  \sum_{j=1}^c \tau_j \prod_{k=1}^r f \left ( x_{k} ; p_{jk}, y_{jk} \right ) ,
\end{align}
where $\tau_j$ is the a priori probability for originating from the $j$'th subpopulation. Thus, the parameters of this mixture model are $\{ y_j \}_{j=1}^c$ with $y_j = (y_{j1}, y_{j2}, \ldots, y_{jr})$ as the central haplotype of the $j$\textsuperscript{th} subpopulation, $\{ \tau_j \}_{j=1}^c$ and $\{ p_{jk} \}_{\substack{j\in \{1, 2, \ldots, c\}\\k \in \{1, 2, \ldots, r\}}}$ (the parameters for each discrete Laplace distribution).

We assume that $p_{jk}$ depends on locus and subpopulation, such that $\log p_{jk} = \omega_j + \lambda_k$. This means that there is an additive effect of locus, $\lambda_k$, and an additive effect of subpopulation, $\omega_j$.

More theory on finite mixture distributions is given by \cite{Titterington1987}.

\subsection{Haplotype frequency prediction}
When we have estimated the parameters of a mixture of multivariate, marginally independent, discrete Laplace distributions (this will be shown in the next section), we can use these to estimate haplotype frequencies.

Given estimates of subpopulation central haplotypes $\{ \hat{y}_{j} \}_{j}$, dispersion parameters $\{ \hat{p}_{jk} \}_{j,k}$ and prior probabilities $\{ \hat{\tau}_{j} \}_{j}$, the haplotype frequency of a haplotype $x = (x_1, x_2, \ldots, x_r)$ with $x_k \in \mathbb{Z}$ for $k \in \{1, 2, \ldots, r\}$ can be estimated as
\begin{align} \label{eq:disclap-predict}
   \hat{p}(x) = \sum_{j=1}^c \hat{\tau}_j \prod_{k=1}^r f \left ( x_k ; \hat{p}_{jk}, \hat{y}_{jk} \right ).
\end{align}

Thus, we simply use the estimated parameters in Equation~\eqref{eq:disclap-mixtures} to obtain Equation~\eqref{eq:disclap-predict}.

\section{Estimating parameters}
In this section we demonstrate how to estimate the parameters in a mixture of multivariate, independent, discrete Laplace distributions. This can for example be used to estimate Y-STR haplotype frequencies.

First, the \texttt{R} package \texttt{disclapmix} \citep{disclapmix12, AndersenDisclap2013} for analysing a mixture of multivariate, independent, discrete Laplace distributions must be loaded:
\begin{knitrout}
\definecolor{shadecolor}{rgb}{0.933, 0.933, 0.933}\color{fgcolor}\begin{kframe}
\begin{alltt}
\hlkwd{library}\hlstd{(disclapmix)}
\end{alltt}
\end{kframe}
\end{knitrout}

If you do not have the \texttt{disclapmix} package installed, please visit \url{http://cran.r-project.org/package=disclapmix}.

This package supplies the function \texttt{disclapmix} for estimating the parameters in a mixture of multivariate, marginally independent, discrete Laplace distributions with probability mass function given in Equation~\eqref{eq:disclap-mixtures}. We will refer to this as 'the discrete Laplace method'.

\subsection{Data from marginally independent, discrete Laplace distributions}
Now, we revisit the example leading to Figure~\ref{fig:simple-example} and add two more loci with different dispersion and location parameters. We then analyse the randomly generated values from independent, discrete Laplace distributions with a probability mass function as given in Equation~\eqref{eq:disclap-independent}.
\begin{knitrout}
\definecolor{shadecolor}{rgb}{0.933, 0.933, 0.933}\color{fgcolor}\begin{kframe}
\begin{alltt}
\hlkwd{set.seed}\hlstd{(}\hlnum{1}\hlstd{)}
\hlstd{n} \hlkwb{<-} \hlnum{100}  \hlcom{# number of individuals}
\hlcom{# Locus 1}
\hlstd{p1} \hlkwb{<-} \hlnum{0.3}  \hlcom{# Dispersion parameter }
\hlstd{m1} \hlkwb{<-} \hlnum{13}  \hlcom{# Location parameter }
\hlstd{d1} \hlkwb{<-} \hlkwd{rdisclap}\hlstd{(n, p1)} \hlopt{+} \hlstd{m1}  \hlcom{# Generate a sample using the rdisclap function}
\hlcom{# Locus 2}
\hlstd{p2} \hlkwb{<-} \hlnum{0.4}
\hlstd{m2} \hlkwb{<-} \hlnum{14}
\hlstd{d2} \hlkwb{<-} \hlkwd{rdisclap}\hlstd{(n, p2)} \hlopt{+} \hlstd{m2}
\hlcom{# Locus 3}
\hlstd{p3} \hlkwb{<-} \hlnum{0.5}
\hlstd{m3} \hlkwb{<-} \hlnum{15}
\hlstd{d3} \hlkwb{<-} \hlkwd{rdisclap}\hlstd{(n, p3)} \hlopt{+} \hlstd{m3}
\hlstd{db} \hlkwb{<-} \hlkwd{cbind}\hlstd{(d1, d2, d3)}
\hlstd{db} \hlkwb{<-} \hlkwd{as.matrix}\hlstd{(}\hlkwd{apply}\hlstd{(db,} \hlnum{2}\hlstd{, as.integer))}  \hlcom{# To coerce to integer matrix}
\hlkwd{head}\hlstd{(db)}
\end{alltt}
\begin{verbatim}
##      d1 d2 d3
## [1,] 14 15 16
## [2,] 12 12 17
## [3,] 13 13 15
## [4,] 13 13 15
## [5,] 14 12 15
## [6,] 13 15 15
\end{verbatim}
\begin{alltt}
\hlcom{# Fit the model (L means integer type)}
\hlstd{fit} \hlkwb{<-} \hlkwd{disclapmix}\hlstd{(db,} \hlkwc{clusters} \hlstd{=} \hlnum{1L}\hlstd{)}
\end{alltt}
\end{kframe}
\end{knitrout}

We can then look at the estimated location parameters, $y=(y_1, y_2, y_3)$:
\begin{knitrout}
\definecolor{shadecolor}{rgb}{0.933, 0.933, 0.933}\color{fgcolor}\begin{kframe}
\begin{alltt}
\hlstd{fit}\hlopt{$}\hlstd{y}
\end{alltt}
\begin{verbatim}
##      d1 d2 d3
## [1,] 13 14 15
\end{verbatim}
\end{kframe}
\end{knitrout}

And the estimated dispersion parameters, $(p_1, p_2, p_3)$:
\begin{knitrout}
\definecolor{shadecolor}{rgb}{0.933, 0.933, 0.933}\color{fgcolor}\begin{kframe}
\begin{alltt}
\hlstd{fit}\hlopt{$}\hlstd{disclap_parameters}
\end{alltt}
\begin{verbatim}
##             d1     d2     d3
## cluster1 0.265 0.4369 0.5167
\end{verbatim}
\end{kframe}
\end{knitrout}

As seen, the estimated dispersion location parameters are well estimated. The dispersion parameters are also quite close to the ones used to generate the data.

\subsection{Data from a Fisher-Wright population}
\cite{AndersenDisclap2013} simulated populations following the Fisher-Wright model of evolution \citep{Fisher1922, Fisher1930, Fisher1958, Wright1931, Ewens2004} with assumptions of primarily neutral, single-step mutations of STRs \citep{OhtaKimura1973}. From these populations, data sets were sampled. Using the discrete Laplace method for estimating haplotype frequencies, the method worked rather well.

This is worth highlighting: Data was simulated under a completely different model than that used for inference afterwards. The data was simulated under a population model (Fisher-Wright model of evolution) with a certain mutation model (single-step mutation model). Inference was made assuming that the data was from a mixture of multivariate, marginally independent, discrete Laplace distributions.

One of the reasons that the discrete Laplace distribution predicts data from a Fisher-Wright model of evolution with a single-step mutation model is due to the fact that it approximates certain properties of this population and mutation model \citep{Caliebe2010}. This is also explained by \cite{AndersenDisclap2013}.

Now, let us try simulating a Fisher-Wright population and analyse it with the discrete Laplace method. To simulate the population, the \texttt{R} package \texttt{fwsim} \citep{fwsim025, fwsim2012Arxiv} is loaded:
\begin{knitrout}
\definecolor{shadecolor}{rgb}{0.933, 0.933, 0.933}\color{fgcolor}\begin{kframe}
\begin{alltt}
\hlkwd{library}\hlstd{(fwsim)}
\end{alltt}
\end{kframe}
\end{knitrout}

If you do not have the \texttt{fwsim} package installed, please visit \url{http://cran.r-project.org/package=fwsim}.

We then simulate a population consisting of Y-STR profiles:
\begin{knitrout}
\definecolor{shadecolor}{rgb}{0.933, 0.933, 0.933}\color{fgcolor}\begin{kframe}
\begin{alltt}
\hlkwd{set.seed}\hlstd{(}\hlnum{1}\hlstd{)}
\hlstd{generations} \hlkwb{<-} \hlnum{100}
\hlstd{population.size} \hlkwb{<-} \hlnum{1e+05}
\hlstd{number.of.loci} \hlkwb{<-} \hlnum{7}
\hlstd{mutation.rates} \hlkwb{<-} \hlkwd{seq}\hlstd{(}\hlnum{0.001}\hlstd{,} \hlnum{0.01}\hlstd{,} \hlkwc{length.out} \hlstd{= number.of.loci)}
\hlstd{mutation.rates}
\end{alltt}
\begin{verbatim}
## [1] 0.0010 0.0025 0.0040 0.0055 0.0070 0.0085 0.0100
\end{verbatim}
\begin{alltt}
\hlstd{sim} \hlkwb{<-} \hlkwd{fwsim}\hlstd{(}\hlkwc{g} \hlstd{= generations,} \hlkwc{k} \hlstd{= population.size,} \hlkwc{r} \hlstd{= number.of.loci,}
    \hlkwc{mu} \hlstd{= mutation.rates,} \hlkwc{trace} \hlstd{=} \hlnum{FALSE}\hlstd{)}
\hlstd{pop} \hlkwb{<-} \hlstd{sim}\hlopt{$}\hlstd{haplotypes}
\end{alltt}
\end{kframe}
\end{knitrout}

Note, that the mutation rates are different for each locus (ranging from 0.001 to 0.01). The location parameter is 0 for all loci by default. This can be changed afterwards without loosing or adding any information. Below, we change it to be $y = (14, 12, 28, 22, 10, 11, 13)$:
\begin{knitrout}
\definecolor{shadecolor}{rgb}{0.933, 0.933, 0.933}\color{fgcolor}\begin{kframe}
\begin{alltt}
\hlstd{y} \hlkwb{<-} \hlkwd{c}\hlstd{(}\hlnum{14}\hlstd{,} \hlnum{12}\hlstd{,} \hlnum{28}\hlstd{,} \hlnum{22}\hlstd{,} \hlnum{10}\hlstd{,} \hlnum{11}\hlstd{,} \hlnum{13}\hlstd{)}
\hlkwa{for} \hlstd{(i} \hlkwa{in} \hlnum{1}\hlopt{:}\hlstd{number.of.loci) \{}
    \hlstd{pop[, i]} \hlkwb{<-} \hlstd{pop[, i]} \hlopt{+} \hlstd{y[i]}
\hlstd{\}}
\hlkwd{head}\hlstd{(pop)}
\end{alltt}
\begin{verbatim}
##   Locus1 Locus2 Locus3 Locus4 Locus5 Locus6 Locus7 N
## 1     12     12     28     22     10     11     13 3
## 2     14     11     26     20      9     11     13 1
## 3     13     11     26     22     10     10     13 4
## 4     14     11     26     22      8     10     13 2
## 5     14     11     26     22      9     10     12 2
## 6     14     11     26     23     10     10     11 2
\end{verbatim}
\end{kframe}
\end{knitrout}

Then, $y$ is the most frequent 10 locus Y-STR haplotype in Denmark according to \url{http://www.yhrd.org} (on March 26, 2013) restricted to the 7 loci minimal haplotype.

The column \texttt{N} is the number of individuals in the population with that Y-STR haplotype. Summing column \texttt{N} reveals that there is not exactly \texttt{population.size} individuals due to that the population size is stochastic (refer to \cite{fwsim2012Arxiv} for the details).

We can then calculate the population frequency for each haplotype:
\begin{knitrout}
\definecolor{shadecolor}{rgb}{0.933, 0.933, 0.933}\color{fgcolor}\begin{kframe}
\begin{alltt}
\hlstd{pop}\hlopt{$}\hlstd{PopFreq} \hlkwb{<-} \hlstd{pop}\hlopt{$}\hlstd{N}\hlopt{/}\hlkwd{sum}\hlstd{(pop}\hlopt{$}\hlstd{N)}
\end{alltt}
\end{kframe}
\end{knitrout}

Let us draw a data set where each haplotype is drawn relatively to its population frequency:
\begin{knitrout}
\definecolor{shadecolor}{rgb}{0.933, 0.933, 0.933}\color{fgcolor}\begin{kframe}
\begin{alltt}
\hlkwd{set.seed}\hlstd{(}\hlnum{1}\hlstd{)}
\hlstd{n} \hlkwb{<-} \hlnum{500}  \hlcom{# Data set size}
\hlstd{types} \hlkwb{<-} \hlkwd{sample}\hlstd{(}\hlkwc{x} \hlstd{=} \hlnum{1}\hlopt{:}\hlkwd{nrow}\hlstd{(pop),} \hlkwc{size} \hlstd{= n,} \hlkwc{replace} \hlstd{=} \hlnum{TRUE}\hlstd{,} \hlkwc{prob} \hlstd{= pop}\hlopt{$}\hlstd{N)}
\hlstd{types.table} \hlkwb{<-} \hlkwd{table}\hlstd{(types)}
\hlstd{alpha} \hlkwb{<-} \hlkwd{sum}\hlstd{(types.table} \hlopt{==} \hlnum{1}\hlstd{)}
\hlstd{alpha}\hlopt{/}\hlstd{n}  \hlcom{# Singleton proportion}
\end{alltt}
\begin{verbatim}
## [1] 0.492
\end{verbatim}
\begin{alltt}
\hlstd{dataset} \hlkwb{<-} \hlstd{pop[}\hlkwd{as.integer}\hlstd{(}\hlkwd{names}\hlstd{(types.table)), ]}
\hlstd{dataset}\hlopt{$}\hlstd{Ndb} \hlkwb{<-} \hlstd{types.table}
\hlkwd{head}\hlstd{(dataset)}
\end{alltt}
\begin{verbatim}
##     Locus1 Locus2 Locus3 Locus4 Locus5 Locus6 Locus7   N   PopFreq Ndb
## 9       14     11     26     23     10      8     12   2 1.924e-05   1
## 103     14     11     28     19      9     10     12   1 9.619e-06   1
## 146     14     11     28     21     10     11     13 187 1.799e-03   3
## 229     14     11     27     21     11     12     12   6 5.771e-05   1
## 271     14     11     28     22      7     11     12  14 1.347e-04   1
## 273     14     11     28     22      8     11     12   6 5.771e-05   1
\end{verbatim}
\begin{alltt}
\hlstd{db} \hlkwb{<-} \hlstd{pop[types,} \hlnum{1}\hlopt{:}\hlstd{number.of.loci]}
\hlstd{db} \hlkwb{<-} \hlkwd{as.matrix}\hlstd{(}\hlkwd{apply}\hlstd{(db,} \hlnum{2}\hlstd{, as.integer))}  \hlcom{# Force it to be an integer matrix}
\hlkwd{head}\hlstd{(db)}
\end{alltt}
\begin{verbatim}
##      Locus1 Locus2 Locus3 Locus4 Locus5 Locus6 Locus7
## [1,]     13     12     30     22      8     11     11
## [2,]     14     12     28     22     10     11     14
## [3,]     14     13     28     21     10     10     14
## [4,]     14     12     28     22      9     11     14
## [5,]     14     12     28     22     11     11     14
## [6,]     14     12     28     22      9     10     14
\end{verbatim}
\end{kframe}
\end{knitrout}

Then, analyse it:
\begin{knitrout}
\definecolor{shadecolor}{rgb}{0.933, 0.933, 0.933}\color{fgcolor}\begin{kframe}
\begin{alltt}
\hlstd{fit} \hlkwb{<-} \hlkwd{disclapmix}\hlstd{(db,} \hlkwc{clusters} \hlstd{=} \hlnum{1L}\hlstd{)}
\hlcom{# Estimated location parameters}
\hlstd{fit}\hlopt{$}\hlstd{y}
\end{alltt}
\begin{verbatim}
##      Locus1 Locus2 Locus3 Locus4 Locus5 Locus6 Locus7
## [1,]     14     12     28     22     10     11     13
\end{verbatim}
\begin{alltt}
\hlcom{# Estimated dispersion parameters}
\hlstd{fit}\hlopt{$}\hlstd{disclap_parameters}
\end{alltt}
\begin{verbatim}
##          Locus1 Locus2 Locus3 Locus4 Locus5 Locus6 Locus7
## cluster1 0.0469  0.126 0.1589 0.1827 0.2453 0.2817  0.316
\end{verbatim}
\end{kframe}
\end{knitrout}

Let us compare the mutation rates with the dispersion parameters in the discrete Laplace distributions:
\begin{figure}[H]
\begin{center}
\begin{knitrout}
\definecolor{shadecolor}{rgb}{0.933, 0.933, 0.933}\color{fgcolor}\begin{kframe}
\begin{alltt}
\hlkwd{plot}\hlstd{(mutation.rates, fit}\hlopt{$}\hlstd{disclap_parameters,} \hlkwc{xlab} \hlstd{=} \hlstr{"Mutation rate"}\hlstd{,}
    \hlkwc{ylab} \hlstd{=} \hlstr{"Estimated dispersion parameter"}\hlstd{)}
\end{alltt}
\end{kframe}
\includegraphics[width=12cm]{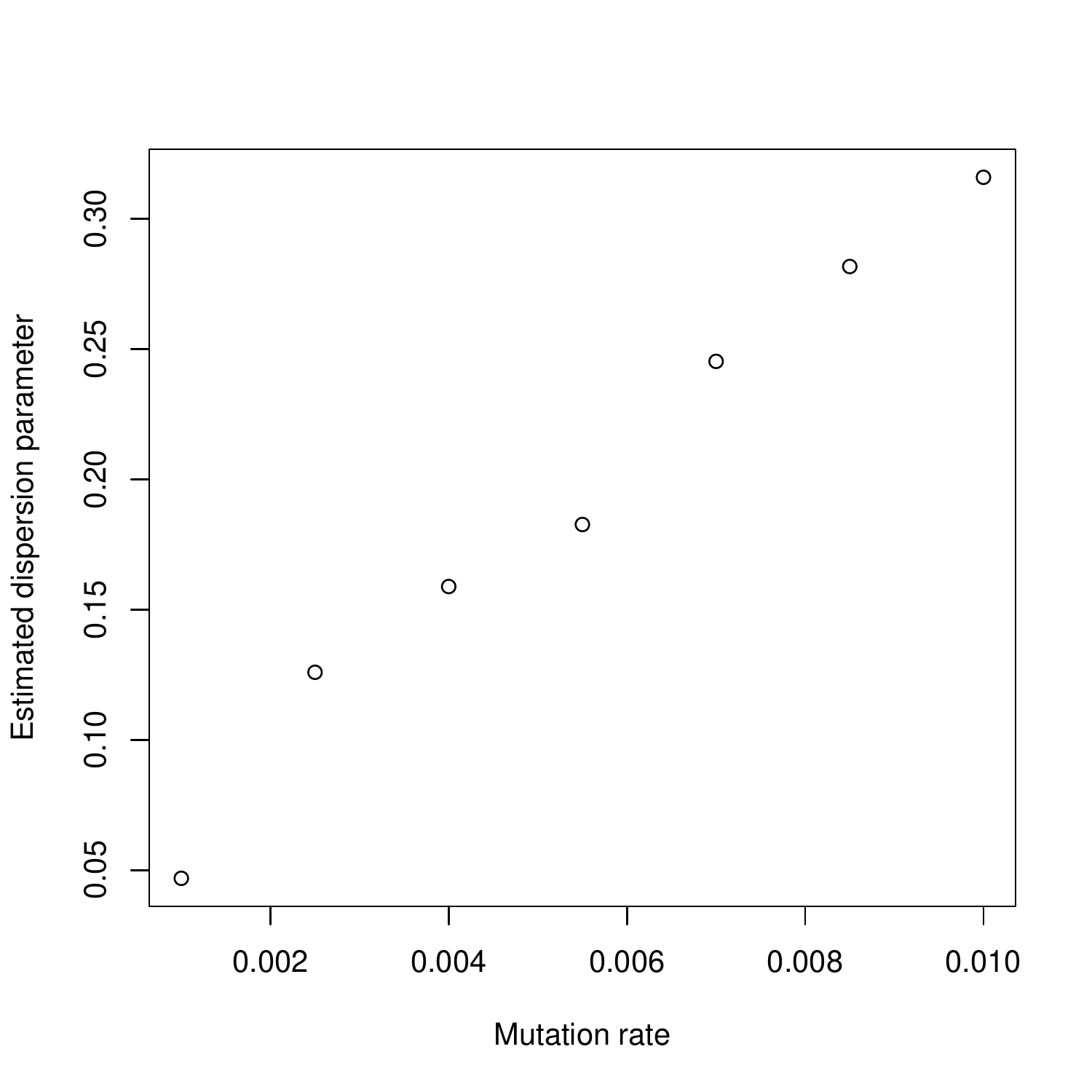} 

\end{knitrout}

\caption{The relationship between the mutation rate in a Fisher-Wright population and the estimated dispersion parameters using the discrete Laplace method.}
\end{center}
\end{figure}
As expected, there is a connection between the mutation rate and the dispersion parameter (the exact connection is not known).

It is possible to predict a population frequency with the \texttt{predict} function as shown in Equation~\eqref{eq:disclap-predict}. This can be used to see how well the population frequency is predicted for each unique haplotype in the dataset (obtained by using \texttt{dataset} instead of \texttt{db}):
\begin{figure}[H]
\begin{center}
\begin{knitrout}
\definecolor{shadecolor}{rgb}{0.933, 0.933, 0.933}\color{fgcolor}\begin{kframe}
\begin{alltt}
\hlstd{pred.popfreqs} \hlkwb{<-} \hlkwd{predict}\hlstd{(fit,}
  \hlkwc{newdata} \hlstd{=} \hlkwd{as.matrix}\hlstd{(}\hlkwd{apply}\hlstd{(dataset[,} \hlnum{1}\hlopt{:}\hlstd{number.of.loci],} \hlnum{2}\hlstd{, as.integer)))}
\hlkwd{plot}\hlstd{(dataset}\hlopt{$}\hlstd{PopFreq, pred.popfreqs,} \hlkwc{log} \hlstd{=} \hlstr{"xy"}\hlstd{,}
  \hlkwc{xlab} \hlstd{=} \hlstr{"True population frequency"}\hlstd{,}
  \hlkwc{ylab} \hlstd{=} \hlstr{"Estimated population frequency"}\hlstd{)}
\hlkwd{abline}\hlstd{(}\hlkwc{a} \hlstd{=} \hlnum{0}\hlstd{,} \hlkwc{b} \hlstd{=} \hlnum{1}\hlstd{,} \hlkwc{lty} \hlstd{=} \hlnum{1}\hlstd{)}
\hlkwd{legend}\hlstd{(}\hlstr{"bottomright"}\hlstd{,} \hlstr{"y = x (predicted = true)"}\hlstd{,} \hlkwc{lty} \hlstd{=} \hlnum{1}\hlstd{)}
\end{alltt}
\end{kframe}
\includegraphics[width=12cm]{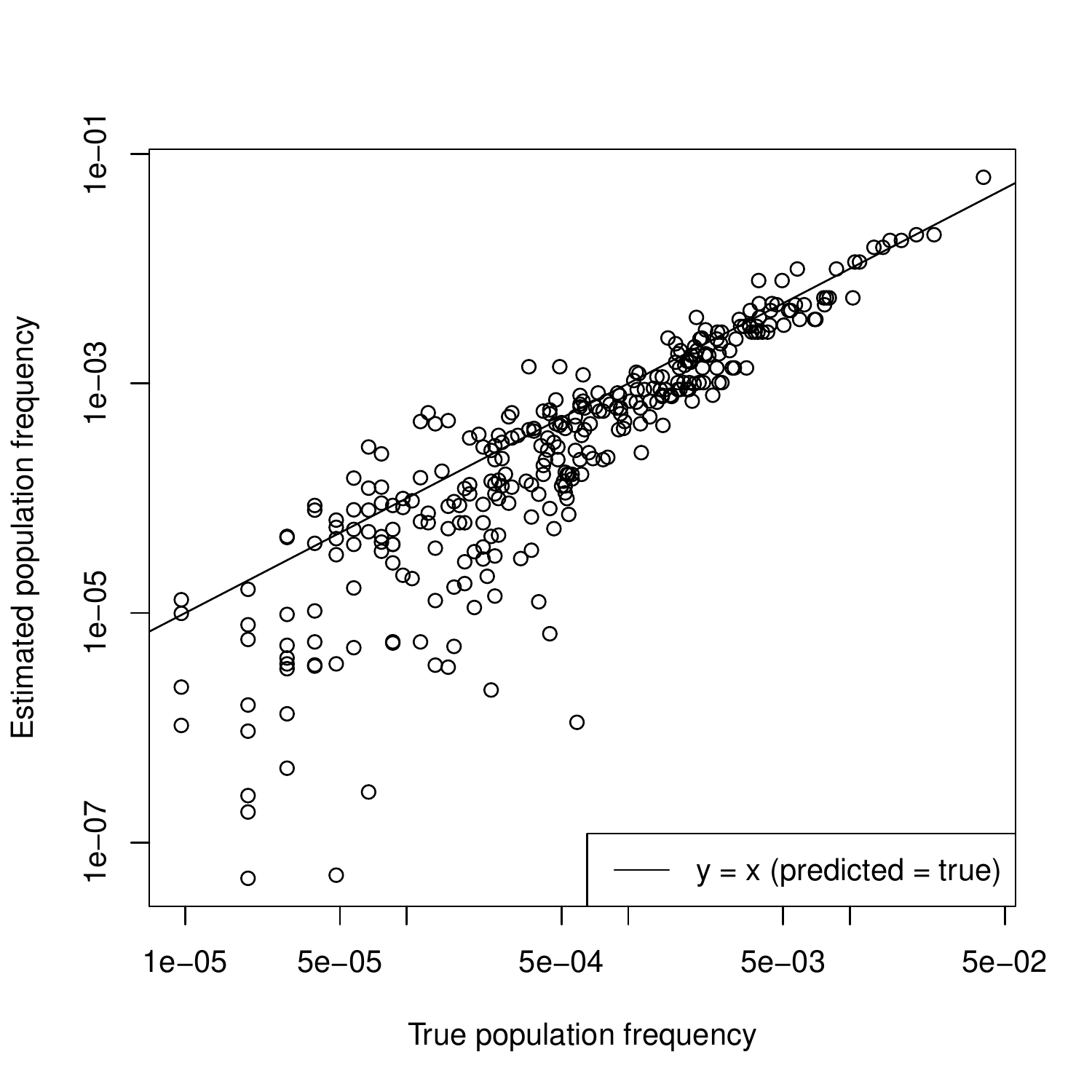} 

\end{knitrout}

\caption{The relationship between the true population frequency and the predicted population frequency using the discrete Laplace method.}
\end{center}
\end{figure}

\subsection{Data from a mixture of two Fisher-Wright populations}
Here, we show how to analyse a dataset from a mixture of two populations. First, we simulate two populations (note the different mutation rates and location parameters, where the location parameters again are changed afterwards without loosing or adding any information):
\begin{knitrout}
\definecolor{shadecolor}{rgb}{0.933, 0.933, 0.933}\color{fgcolor}\begin{kframe}
\begin{alltt}
\hlkwd{set.seed}\hlstd{(}\hlnum{1}\hlstd{)}
\hlcom{# Common parameters}
\hlstd{generations} \hlkwb{<-} \hlnum{100}
\hlstd{population.size} \hlkwb{<-} \hlnum{1e+05}
\hlstd{number.of.loci} \hlkwb{<-} \hlnum{7}
\hlstd{mu1} \hlkwb{<-} \hlkwd{seq}\hlstd{(}\hlnum{0.001}\hlstd{,} \hlnum{0.005}\hlstd{,} \hlkwc{length.out} \hlstd{= number.of.loci)}
\hlstd{sim1} \hlkwb{<-} \hlkwd{fwsim}\hlstd{(}\hlkwc{g} \hlstd{= generations,} \hlkwc{k} \hlstd{= population.size,} \hlkwc{r} \hlstd{= number.of.loci,}
    \hlkwc{mu} \hlstd{= mu1,} \hlkwc{trace} \hlstd{=} \hlnum{FALSE}\hlstd{)}
\hlstd{pop1} \hlkwb{<-} \hlstd{sim1}\hlopt{$}\hlstd{haplotypes}
\hlstd{y1} \hlkwb{<-} \hlkwd{c}\hlstd{(}\hlnum{14}\hlstd{,} \hlnum{12}\hlstd{,} \hlnum{28}\hlstd{,} \hlnum{22}\hlstd{,} \hlnum{10}\hlstd{,} \hlnum{11}\hlstd{,} \hlnum{13}\hlstd{)}
\hlkwa{for} \hlstd{(i} \hlkwa{in} \hlnum{1}\hlopt{:}\hlstd{number.of.loci) pop1[, i]} \hlkwb{<-} \hlstd{pop1[, i]} \hlopt{+} \hlstd{y1[i]}
\hlstd{mu2} \hlkwb{<-} \hlkwd{seq}\hlstd{(}\hlnum{0.005}\hlstd{,} \hlnum{0.01}\hlstd{,} \hlkwc{length.out} \hlstd{= number.of.loci)}
\hlstd{sim2} \hlkwb{<-} \hlkwd{fwsim}\hlstd{(}\hlkwc{g} \hlstd{= generations,} \hlkwc{k} \hlstd{= population.size,} \hlkwc{r} \hlstd{= number.of.loci,}
    \hlkwc{mu} \hlstd{= mu2,} \hlkwc{trace} \hlstd{=} \hlnum{FALSE}\hlstd{)}
\hlstd{pop2} \hlkwb{<-} \hlstd{sim2}\hlopt{$}\hlstd{haplotypes}
\hlstd{y2} \hlkwb{<-} \hlkwd{c}\hlstd{(}\hlnum{14}\hlstd{,} \hlnum{13}\hlstd{,} \hlnum{29}\hlstd{,} \hlnum{23}\hlstd{,} \hlnum{11}\hlstd{,} \hlnum{13}\hlstd{,} \hlnum{13}\hlstd{)}
\hlkwa{for} \hlstd{(i} \hlkwa{in} \hlnum{1}\hlopt{:}\hlstd{number.of.loci) pop2[, i]} \hlkwb{<-} \hlstd{pop2[, i]} \hlopt{+} \hlstd{y2[i]}
\end{alltt}
\end{kframe}
\end{knitrout}

Here, just as $y_1 = (14, 12, 28, 22, 10, 11, 13)$ are the alleles from most frequent haplotype, then $y_2 = (14, 13, 29, 23, 11, 13, 13)$ are the alleles from the second most frequent haplotype.

Then we sample a data set with an expected proportion of 20\% from the first population and 80\% from the second population:
\begin{knitrout}
\definecolor{shadecolor}{rgb}{0.933, 0.933, 0.933}\color{fgcolor}\begin{kframe}
\begin{alltt}
\hlkwd{set.seed}\hlstd{(}\hlnum{1}\hlstd{)}
\hlstd{n} \hlkwb{<-} \hlnum{500}  \hlcom{# Data set size}
\hlstd{n1} \hlkwb{<-} \hlkwd{rbinom}\hlstd{(}\hlnum{1}\hlstd{, n,} \hlnum{0.2}\hlstd{)}
\hlkwd{c}\hlstd{(n1, n1}\hlopt{/}\hlstd{n)}
\end{alltt}
\begin{verbatim}
## [1] 102.000   0.204
\end{verbatim}
\begin{alltt}
\hlstd{n2} \hlkwb{<-} \hlstd{n} \hlopt{-} \hlstd{n1}
\hlkwd{c}\hlstd{(n2, n2}\hlopt{/}\hlstd{n)}
\end{alltt}
\begin{verbatim}
## [1] 398.000   0.796
\end{verbatim}
\begin{alltt}
\hlstd{types1} \hlkwb{<-} \hlkwd{sample}\hlstd{(}\hlkwc{x} \hlstd{=} \hlnum{1}\hlopt{:}\hlkwd{nrow}\hlstd{(pop1),} \hlkwc{size} \hlstd{= n1,} \hlkwc{replace} \hlstd{=} \hlnum{TRUE}\hlstd{,} \hlkwc{prob} \hlstd{= pop1}\hlopt{$}\hlstd{N)}
\hlstd{db1} \hlkwb{<-} \hlstd{pop1[types1,} \hlnum{1}\hlopt{:}\hlstd{number.of.loci]}
\hlstd{types2} \hlkwb{<-} \hlkwd{sample}\hlstd{(}\hlkwc{x} \hlstd{=} \hlnum{1}\hlopt{:}\hlkwd{nrow}\hlstd{(pop2),} \hlkwc{size} \hlstd{= n2,} \hlkwc{replace} \hlstd{=} \hlnum{TRUE}\hlstd{,} \hlkwc{prob} \hlstd{= pop2}\hlopt{$}\hlstd{N)}
\hlstd{db2} \hlkwb{<-} \hlstd{pop2[types2,} \hlnum{1}\hlopt{:}\hlstd{number.of.loci]}
\hlstd{db} \hlkwb{<-} \hlkwd{rbind}\hlstd{(db1, db2)}
\hlstd{db} \hlkwb{<-} \hlkwd{as.matrix}\hlstd{(}\hlkwd{apply}\hlstd{(db,} \hlnum{2}\hlstd{, as.integer))}  \hlcom{# Force it to be an integer matrix}
\hlcom{# Singleton proportion}
\hlkwd{sum}\hlstd{(}\hlkwd{table}\hlstd{(}\hlkwd{apply}\hlstd{(db,} \hlnum{1}\hlstd{, paste,} \hlkwc{collapse} \hlstd{=} \hlstr{";"}\hlstd{))} \hlopt{==} \hlnum{1}\hlstd{)}\hlopt{/}\hlstd{n}
\end{alltt}
\begin{verbatim}
## [1] 0.672
\end{verbatim}
\end{kframe}
\end{knitrout}

Now, we analyse the data set trying 1 to 5 subpopulations. Afterwards, we analyse the optimal number of subpopulations using the BIC (Bayesian Information Criteria) by \cite{BIC}:
\begin{knitrout}
\definecolor{shadecolor}{rgb}{0.933, 0.933, 0.933}\color{fgcolor}\begin{kframe}
\begin{alltt}
\hlstd{fits} \hlkwb{<-} \hlkwd{lapply}\hlstd{(}\hlnum{1L}\hlopt{:}\hlnum{5L}\hlstd{,} \hlkwa{function}\hlstd{(}\hlkwc{clusters}\hlstd{)} \hlkwd{disclapmix}\hlstd{(db,} \hlkwc{clusters} \hlstd{= clusters))}
\end{alltt}
\end{kframe}
\end{knitrout}

The BIC values are:
\begin{knitrout}
\definecolor{shadecolor}{rgb}{0.933, 0.933, 0.933}\color{fgcolor}\begin{kframe}
\begin{alltt}
\hlstd{BIC} \hlkwb{<-} \hlkwd{sapply}\hlstd{(fits,} \hlkwa{function}\hlstd{(}\hlkwc{fit}\hlstd{) fit}\hlopt{$}\hlstd{BIC_marginal)}
\hlstd{BIC}
\end{alltt}
\begin{verbatim}
## [1] 9487 8600 8646 8700 8748
\end{verbatim}
\end{kframe}
\end{knitrout}

The estimated parameters for this optimal number of subpopulations can be made available in \texttt{best.fit} as follows:
\begin{knitrout}
\definecolor{shadecolor}{rgb}{0.933, 0.933, 0.933}\color{fgcolor}\begin{kframe}
\begin{alltt}
\hlstd{best.fit} \hlkwb{<-} \hlstd{fits[[}\hlkwd{which.min}\hlstd{(BIC)]]}
\hlstd{best.fit}
\end{alltt}
\begin{verbatim}
## disclapmixfit from 500 observations on 7 loci with 2 clusters.
\end{verbatim}
\begin{alltt}
\hlcom{# Estimated a priori probability of originating from each}
\hlcom{# subpopulation}
\hlstd{best.fit}\hlopt{$}\hlstd{tau}
\end{alltt}
\begin{verbatim}
## [1] 0.2126 0.7874
\end{verbatim}
\begin{alltt}
\hlcom{# Estimated location parameters}
\hlstd{best.fit}\hlopt{$}\hlstd{y}
\end{alltt}
\begin{verbatim}
##      Locus1 Locus2 Locus3 Locus4 Locus5 Locus6 Locus7
## [1,]     14     12     28     22     10     11     13
## [2,]     14     13     29     23     11     13     13
\end{verbatim}
\begin{alltt}
\hlcom{# Estimated dispersion parameters for each subpopulation}
\hlstd{best.fit}\hlopt{$}\hlstd{disclap_parameters}
\end{alltt}
\begin{verbatim}
##          Locus1 Locus2 Locus3 Locus4 Locus5 Locus6 Locus7
## cluster1 0.1029 0.1083 0.1213 0.1353 0.1458 0.1587 0.1595
## cluster2 0.1896 0.1997 0.2234 0.2494 0.2686 0.2924 0.2938
\end{verbatim}
\end{kframe}
\end{knitrout}

The estimated location parameters are the same as those used for generating the data. Also, the values of $\tau_j$, the a priori probability of originating from the $j$\textsuperscript{th} subpopulation, are consistent with the mixture proportions of 0.204 and 0.796.

We can also calculate the predicted population frequencies (using the mixture proportions 0.204 and 0.796):
\begin{knitrout}
\definecolor{shadecolor}{rgb}{0.933, 0.933, 0.933}\color{fgcolor}\begin{kframe}
\begin{alltt}
\hlstd{pop1}\hlopt{$}\hlstd{PopFreq} \hlkwb{<-} \hlstd{pop1}\hlopt{$}\hlstd{N}\hlopt{/}\hlkwd{sum}\hlstd{(pop1}\hlopt{$}\hlstd{N)}
\hlstd{pop2}\hlopt{$}\hlstd{PopFreq} \hlkwb{<-} \hlstd{pop2}\hlopt{$}\hlstd{N}\hlopt{/}\hlkwd{sum}\hlstd{(pop2}\hlopt{$}\hlstd{N)}
\hlstd{types1.table} \hlkwb{<-} \hlkwd{table}\hlstd{(types1)}
\hlstd{types2.table} \hlkwb{<-} \hlkwd{table}\hlstd{(types2)}
\hlstd{dataset1} \hlkwb{<-} \hlstd{pop1[}\hlkwd{as.integer}\hlstd{(}\hlkwd{names}\hlstd{(types1.table)), ]}
\hlstd{dataset1}\hlopt{$}\hlstd{Ndb} \hlkwb{<-} \hlstd{types1.table}
\hlkwd{sum}\hlstd{(dataset1}\hlopt{$}\hlstd{Ndb)}
\end{alltt}
\begin{verbatim}
## [1] 102
\end{verbatim}
\begin{alltt}
\hlstd{dataset2} \hlkwb{<-} \hlstd{pop2[}\hlkwd{as.integer}\hlstd{(}\hlkwd{names}\hlstd{(types2.table)), ]}
\hlstd{dataset2}\hlopt{$}\hlstd{Ndb} \hlkwb{<-} \hlstd{types2.table}
\hlkwd{sum}\hlstd{(dataset2}\hlopt{$}\hlstd{Ndb)}
\end{alltt}
\begin{verbatim}
## [1] 398
\end{verbatim}
\begin{alltt}
\hlstd{dataset} \hlkwb{<-} \hlkwd{merge}\hlstd{(}\hlkwc{x} \hlstd{= dataset1,} \hlkwc{y} \hlstd{= dataset2,} \hlkwc{by} \hlstd{=} \hlkwd{colnames}\hlstd{(db),} \hlkwc{all} \hlstd{=} \hlnum{TRUE}\hlstd{)}
\hlstd{dataset[}\hlkwd{is.na}\hlstd{(dataset)]} \hlkwb{<-} \hlnum{0}
\hlstd{dataset}\hlopt{$}\hlstd{MixPopFreq} \hlkwb{<-} \hlstd{(n1}\hlopt{/}\hlstd{n)} \hlopt{*} \hlstd{dataset}\hlopt{$}\hlstd{PopFreq.x} \hlopt{+} \hlstd{(n2}\hlopt{/}\hlstd{n)} \hlopt{*} \hlstd{dataset}\hlopt{$}\hlstd{PopFreq.y}
\hlstd{dataset}\hlopt{$}\hlstd{Type} \hlkwb{<-} \hlstr{"Only from pop1"}
\hlstd{dataset}\hlopt{$}\hlstd{Type[dataset}\hlopt{$}\hlstd{Ndb.y} \hlopt{>} \hlnum{0}\hlstd{]} \hlkwb{<-} \hlstr{"Only from pop2"}
\hlstd{dataset}\hlopt{$}\hlstd{Type[dataset}\hlopt{$}\hlstd{Ndb.x} \hlopt{>} \hlnum{0} \hlopt{&} \hlstd{dataset}\hlopt{$}\hlstd{Ndb.y} \hlopt{>} \hlnum{0}\hlstd{]} \hlkwb{<-} \hlstr{"Occurred in both"}
\hlstd{dataset}\hlopt{$}\hlstd{Type} \hlkwb{<-} \hlkwd{factor}\hlstd{(dataset}\hlopt{$}\hlstd{Type)}
\end{alltt}
\end{kframe}
\end{knitrout}

\newpage
We can now compare the predicted frequencies with the population frequency:
\begin{figure}[H]
\begin{center}
\begin{knitrout}
\definecolor{shadecolor}{rgb}{0.933, 0.933, 0.933}\color{fgcolor}\begin{kframe}
\begin{alltt}
\hlstd{pred.popfreqs} \hlkwb{<-} \hlkwd{predict}\hlstd{(best.fit,}
  \hlkwc{newdata} \hlstd{=} \hlkwd{as.matrix}\hlstd{(}\hlkwd{apply}\hlstd{(dataset[,} \hlnum{1}\hlopt{:}\hlstd{number.of.loci],} \hlnum{2}\hlstd{, as.integer)))}
\hlkwd{plot}\hlstd{(dataset}\hlopt{$}\hlstd{MixPopFreq, pred.popfreqs,} \hlkwc{log} \hlstd{=} \hlstr{"xy"}\hlstd{,} \hlkwc{col} \hlstd{= dataset}\hlopt{$}\hlstd{Type,}
  \hlkwc{xlab} \hlstd{=} \hlstr{"True population frequency"}\hlstd{,}
  \hlkwc{ylab} \hlstd{=} \hlstr{"Estimated population frequency"}\hlstd{)}
\hlkwd{abline}\hlstd{(}\hlkwc{a} \hlstd{=} \hlnum{0}\hlstd{,} \hlkwc{b} \hlstd{=} \hlnum{1}\hlstd{,} \hlkwc{lty} \hlstd{=} \hlnum{1}\hlstd{)}
\hlkwd{legend}\hlstd{(}\hlstr{"bottomright"}\hlstd{,} \hlkwd{c}\hlstd{(}\hlstr{"y = x (predicted = true)"}\hlstd{,} \hlkwd{levels}\hlstd{(dataset}\hlopt{$}\hlstd{Type)),}
  \hlkwc{lty} \hlstd{=} \hlkwd{c}\hlstd{(}\hlnum{1}\hlstd{,} \hlkwd{rep}\hlstd{(}\hlopt{-}\hlnum{1}\hlstd{,} \hlnum{3}\hlstd{)),} \hlkwc{col} \hlstd{=} \hlkwd{c}\hlstd{(}\hlstr{"black"}\hlstd{,} \hlnum{1}\hlopt{:}\hlkwd{length}\hlstd{(}\hlkwd{levels}\hlstd{(dataset}\hlopt{$}\hlstd{Type))),}
  \hlkwc{pch} \hlstd{=} \hlkwd{c}\hlstd{(}\hlopt{-}\hlnum{1}\hlstd{,} \hlkwd{rep}\hlstd{(}\hlnum{1}\hlstd{,} \hlnum{3}\hlstd{)))}
\end{alltt}
\end{kframe}
\includegraphics[width=12cm]{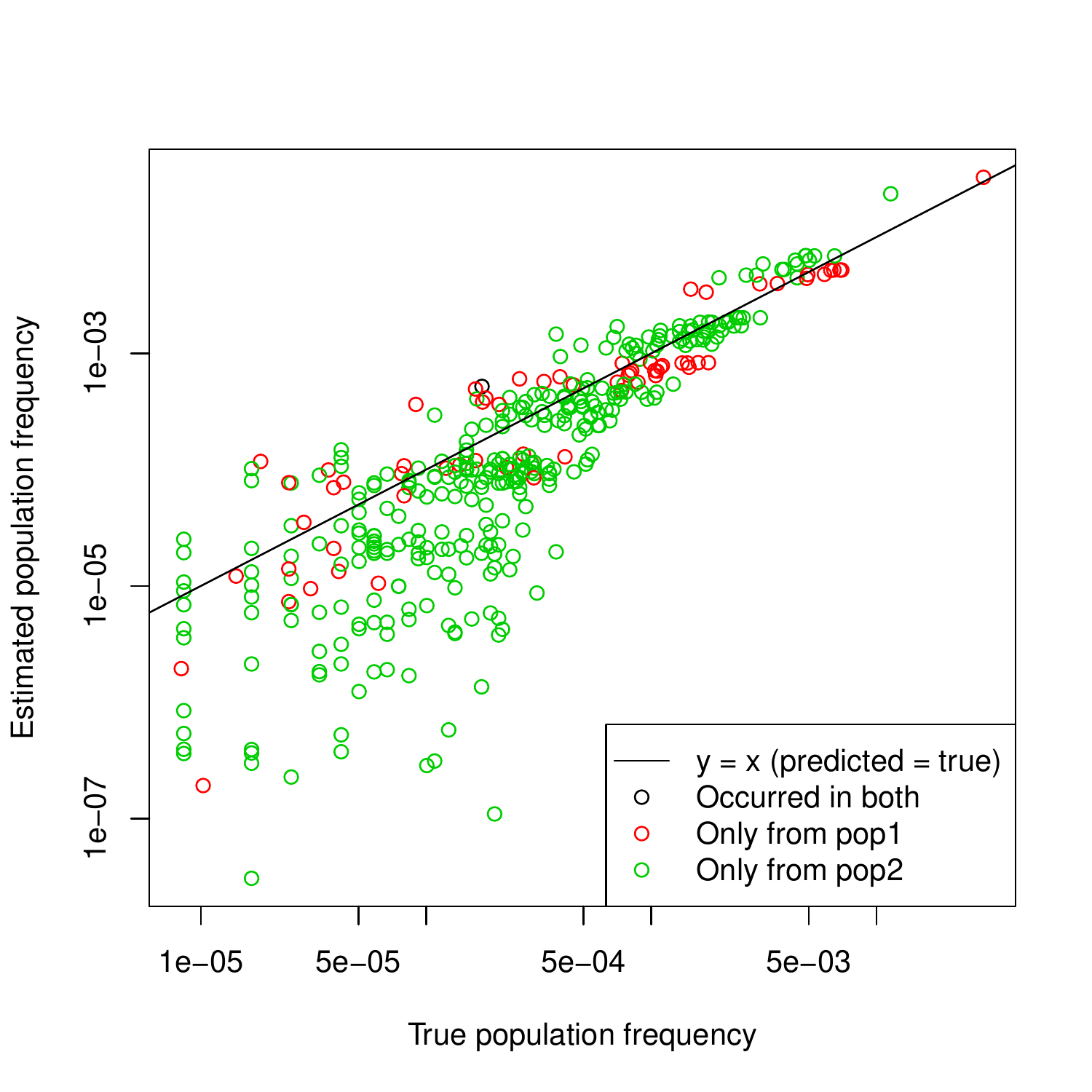} 

\end{knitrout}

\caption{The relationship between the true population frequency and the predicted population frequency using the discrete Laplace method.}
\end{center}
\end{figure}

\newpage
\section{Concluding remarks}
We have shown how to analyse Y-STR population data using the discrete Laplace method described by \cite{AndersenDisclap2013}. This was done using the freely available and open-source \texttt{R} packages \texttt{disclap}, \texttt{fwsim} and \texttt{disclapmix} that are supported on Linux, MacOS and MS Windows.

One key point made is worth repeating: Data simulated under a population model (e.g.\ the Fisher-Wright model of evolution) with a certain mutation model (e.g.\ the single-step mutation model) can be successfully analysed using the discrete Laplace method making inference assuming that the data is from a mixture of multivariate, independent, discrete Laplace distributions.

\bibliographystyle{plainnat}

\begin{thebibliography}{15}
\providecommand{\natexlab}[1]{#1}
\providecommand{\url}[1]{\texttt{#1}}
\expandafter\ifx\csname urlstyle\endcsname\relax
  \providecommand{\doi}[1]{doi: #1}\else
  \providecommand{\doi}{doi: \begingroup \urlstyle{rm}\Url}\fi

\bibitem[Andersen and Eriksen(2012{\natexlab{a}})]{fwsim025}
Mikkel~Meyer Andersen and Poul~Svante Eriksen.
\newblock \emph{fwsim: Fisher-Wright Population Simulation},
  2012{\natexlab{a}}.
\newblock URL \url{http://CRAN.R-project.org/package=fwsim}.
\newblock R~package version~0.2-5.

\bibitem[Andersen and Eriksen(2012{\natexlab{b}})]{fwsim2012Arxiv}
Mikkel~Meyer Andersen and Poul~Svante Eriksen.
\newblock Efficient forward simulation of fisher-wright populations with
  stochastic population size and neutral single step mutations in haplotypes.
\newblock \emph{Preprint}, 2012{\natexlab{b}}.
\newblock arXiv:1210.1773.

\bibitem[Andersen and Eriksen(2013{\natexlab{a}})]{disclap14}
Mikkel~Meyer Andersen and Poul~Svante Eriksen.
\newblock \emph{disclap: Discrete Laplace Family}, 2013{\natexlab{a}}.
\newblock URL \url{http://CRAN.R-project.org/package=disclap}.
\newblock R package version 1.4.

\bibitem[Andersen and Eriksen(2013{\natexlab{b}})]{disclapmix12}
Mikkel~Meyer Andersen and Poul~Svante Eriksen.
\newblock \emph{disclapmix: Discrete Laplace mixture inference using the EM
  algorithm}, 2013{\natexlab{b}}.
\newblock URL \url{http://CRAN.R-project.org/package=disclapmix}.
\newblock R package version 1.2.

\bibitem[Andersen et~al.(2013)Andersen, Eriksen, and
  Morling]{AndersenDisclap2013}
Mikkel~Meyer Andersen, Poul~Svante Eriksen, and Niels Morling.
\newblock {The discrete Laplace exponential family and estimation of Y-STR
  haplotype frequencies}.
\newblock \emph{Journal of Theoretical Biology}, 2013.
\newblock In press: \url{http://dx.doi.org/10.1016/j.jtbi.2013.03.009}.

\bibitem[Caliebe et~al.(2010)Caliebe, Jochens, Krawczak, and
  R{\"o}sler]{Caliebe2010}
Amke Caliebe, Arne Jochens, Michael Krawczak, and Uwe R{\"o}sler.
\newblock {A Markov Chain Description of the Stepwise Mutation Model: Local and
  Global Behaviour of the Allele Process}.
\newblock \emph{Journal of Theoretical Biology}, 266\penalty0 (2):\penalty0
  336--342, 2010.
\newblock ISSN 0022-5193.

\bibitem[Ewens(2004)]{Ewens2004}
Warren~J. Ewens.
\newblock \emph{Mathematical Population Genetics}.
\newblock Springer-Verlag, 2004.

\bibitem[Fisher(1922)]{Fisher1922}
R.~A. Fisher.
\newblock {On the Dominance Ratio}.
\newblock \emph{Proc. Roy. Soc. Edin.}, 42:\penalty0 321--341, 1922.

\bibitem[Fisher(1930)]{Fisher1930}
R.~A. Fisher.
\newblock \emph{The Genetical Theory of Natural Selection}.
\newblock Oxford: Clarendon Press, 1930.

\bibitem[Fisher(1958)]{Fisher1958}
R.~A. Fisher.
\newblock \emph{The Genetical Theory of Natural Selection}.
\newblock New York: Dover, 2nd revised edition, 1958.

\bibitem[Ohta and Kimura(1973)]{OhtaKimura1973}
T.~Ohta and M.~Kimura.
\newblock {A Model of Mutation Appropriate to Estimate the Number of
  Electrophoretically Detectable Alleles in a Finite Population}.
\newblock \emph{Genet. Res.}, 22:\penalty0 201--204, 1973.

\bibitem[{R Development Core Team}(2012)]{R}
{R Development Core Team}.
\newblock \emph{{R: A Language and Environment for Statistical Computing}}.
\newblock R Foundation for Statistical Computing, Vienna, Austria, 2012.
\newblock URL \url{http://www.R-project.org}.
\newblock {ISBN} 3-900051-07-0.

\bibitem[Schwarz(1978)]{BIC}
Gideon Schwarz.
\newblock {Estimating the Dimension of a Model}.
\newblock \emph{Annals of Statistics}, 6\penalty0 (2):\penalty0 461--464, 1978.

\bibitem[Titterington et~al.(1987)Titterington, Smith, and
  Makov]{Titterington1987}
D.~M. Titterington, A.~F.~M. Smith, and U.~E. Makov.
\newblock \emph{{Statistical Analysis of Finite Mixture Distributions}}.
\newblock Wiley, 1987.

\bibitem[Wright(1931)]{Wright1931}
S.~Wright.
\newblock {Evolution in Mendelian populations}.
\newblock \emph{Genetics}, 16:\penalty0 97--159, 1931.

\end{thebibliography}

\end{document}